# ON THE RADIATIVE AND THERMODYNAMIC PROPERTIES OF THE COSMIC MICROWAVE BACKGROUND RADIATION USING *COBE* FIRAS INSTRUMENT DATA

**Anatoliy I Fisenko, Vladimir Lemberg**

*ONCFEC Inc., 250 Lake Street, Suite 909, St. Catharines, Ontario L2R 5Z4, Canada*

## ABSTRACT

Use formulas to describe the monopole and dipole spectra of the Cosmic Microwave Background (CMB) radiation, the exact expressions for the temperature dependences of the radiative and thermodynamic functions, such as the total radiation power per unit area, total energy density, number density of photons, Helmholtz free energy density, entropy density, heat capacity at constant volume, pressure, enthalpy density, and internal energy density in the finite range of frequencies $\nu_1 \leq \nu \leq \nu_2$ are obtained. Since the dependence of temperature upon the redshift $z$ is known, the obtained expressions can be simply presented in $z$ representation. Utilizing experimental data for the monopole and dipole spectra measured by the *COBE* FIRAS instrument in the 60 - 600 GHz frequency interval at the temperature $T$ = 2.728 K, the values of the radiative and thermodynamic functions, as well as the radiation density constant $a$ and the Stefan-Boltzmann constant σ are calculated. In the case of the dipole spectrum, the constants $a$ and σ, and the radiative and thermodynamic properties of the CMB radiation are obtained using the mean amplitude $T_{amp}$ =3.369m K. It is shown that the Doppler shift leads to a renormalization of the radiation density constant $a$, the Stefan-Boltzmann constant σ, and the corresponding constants for the thermodynamic functions. The radiative and thermodynamic properties of the Cosmic Microwave Background radiation for the monopole and dipole spectra at the redshift z ≈ 1089 are calculated.

## 1. INTRODUCTION

In many areas of astrophysics and physics associated with the Far InfraRed Absolute Spectrophotometry (FIRAS), Fourier Transform InfraRed (FTIR) spectroscopy, radiation pyrometry, or any other area of study of electromagnetic radiation, the spectrum of a real body in the finite frequency interval needs to be measured.

In FIRAS, for example, only a small region of the electromagnetic spectrum is of present concern; 2 - 20 $cm^{-1}$ frequency interval for the measurement of the Cosmic Microwave Background (CMB) radiation (Mather et al. 1990; Fixsen et al. 1994) or $5 - 80\,cm^{-1}$ frequency interval to observe the spectrum of the extragalactic Far InfraRed Background radiation (Fixen et al. 1998). This instrument has been developed to determine the spectral radiation intensity seen through these short radiation windows. The theoretical experiments utilizing a computer require the calculation of the total radiation, number density of photons, as well as the thermodynamic functions, as seen through these short windows. A numerical solution for the computer calculation of these functions should be obtained over a specified range of the spectrum. The present work is devoted to the construction of the exact expressions for the radiative and thermodynamic functions of the CMB radiation in the finite range of frequencies, which can be used for the computer calculation.

It is well-known that the discovery of the Cosmic Microwave Background (CMB) radiation by Penzias & Wilson (1965) provides a strong observational foundation for confirming the Gamow's primordial-fireball hypothesis (Sunyaev & Zel'dovich, 1980). The theoretical prediction that the CMB spectrum is so close to a blackbody was confirmed by the *COBE* FIRAS observation (Mather et al. 1990). However, the perfect fitting of the measured spectrum with the spectrum of a blackbody was achieved only at the peak of the blackbody in the 2 - 20 $cm^{-1}$ frequency interval. This range belongs to the Planck part of the total spectrum. In the Rayleigh-Jeans approximation at low frequency, most experiments are consistent with $T$ = 2.735 ± 0.06 K, but some are not. With respect to the Wien part of the total spectrum, the spectral distortions of the spectrum are difficult to measure due to the foreground signal from interstellar dust at high frequencies (Gawiser & Silk, 2000). Above a few hundred GHz, the detected an isotropic Far-InfraRed Background dominates the Cosmic Microwave Background (Fixen et al. 1998; Puget et al.1996; Dwek et al. 1998; Schlegel et al. 1998; Burigana & Popa 1998).

As a result, the theoretical prediction that the CMB spectrum has a blackbody one in the semi-infinite range of frequencies should be confirmed by new experiments, especially in Wien's part of the spectrum.

In order to construct the thermodynamics of the CMB radiation, knowledge of integral characteristics of a system, such as the total energy density and the number density of photons, is necessary. In this case, the experimental data of the CMB radiation spectrum in a wide range of frequencies should be used. With regards to the Rayleigh-Jeans part of the spectrum, the measurements of the spectral energy density were obtained on several frequencies only by different researches. Thus, their data is not sufficient for constructing the thermodynamics of the CMB radiation, as the data were obtained at different times using different observing strategies, with different devices of varying accuracy. Currently, only the data measured by the *COBE* FIRAS instrument within a finite range of frequencies, ranging from 2 $cm^{-1}$ to 20 $cm^{-1}$, which cover the Planck region, is reliable for the calculation of the thermodynamic functions of the CMB radiation.

In previous studies (Fisenko & Ivashov 2009; Fisenko & Lemberg 2012; Fisenko & Lemberg 2013), the thermodynamics of the thermal radiation of real bodies, such as molybdenium, luminous flames, stoichiometric carbides of hafnium, titanium and zirconium, and $ZrB_2$-SiC-based ultra-high temperature ceramics in the finite range of frequencies at high temperatures were constructed. The calculated values of radiative and thermodynamic functions were in good agreement with experimental data.

In the present work, the exact expressions for the temperature dependences of the radiative and thermodynamic functions, such as the total radiation power per unit area, total energy density, radiation density constant $a$, Stefan-Boltzmann's constant $\sigma$, number density of photons, Helmholtz free energy density, entropy density, heat capacity at constant volume, pressure, enthalpy density, and internal energy density for the monopole and dipole spectra in the finite range of frequencies are constructed. These results can be presented in the redshift z representation. For the monopole spectrum, the values of the radiative and thermodynamic functions at the mean temperature $T$ = 2.728 K are calculated. In the case of the dipole spectrum, the calculated values of the radiative and thermodynamic functions were obtained using the mean amplitude $T_{\rm amp} = 3.369$ mK . It is shown that the Doppler shift leads to a renormalization of the radiation density constant, the Stefan-Boltzmann constant, and the corresponding constants for the thermodynamic functions. The radiative and thermodynamic functions of the CMB radiation for the monopole and dipole spectra at the redshift $z \approx 1089$ are presented.

## 2. GENERAL RELATIONSHIPS FOR MONOPOLE AND DIPOLE SPECTRA

According to (Mather et al. 1994), the cosmological anisotropy is predicted to have a Planckian spectrum of the following form

$$I_0(\tilde{\nu}) \approx B_{\tilde{\nu}}(T_0) + \frac{\partial B_{\tilde{\nu}}(T)}{\partial T}\Big|_{T=T_0} \Delta T \quad , \tag{1}$$

Here the two terms (monopole and dipole) are the Planck blackbody radiation spectrum with the temperature $T_0 + \Delta T$ . $T_0 = 2.728$ K is the mean temperature of the CMB radiation (Fixsen et al. 1996). The temperature fluctuation $\Delta T = T - T_0$ is the temperature anisotropy in a given direction in the sky and can be presented in the form $\Delta T \approx \left(\frac{\upsilon}{c}\right) T_{\mathrm{amp}} \cos(\theta)$. Here $\upsilon$ is a velocity of moving observer with respect to the rest frame of the blackbody radiation, $\theta$ is the angle between the direction of observation and the dipole direction $(l,b) = (264.^{\mathrm{o}}26, +48.^{\mathrm{o}}22)$ (Bennett et al. 1996), and the dipole amplitude $T_{\mathrm{amp}} = 3.369 \pm 0.004$ mK . $B_{\tilde{\nu}}(T)$ at the temperature $T$ is given by the Planck law

$$B_{\tilde{\nu}}(T) = 2hc^2 \frac{\tilde{\nu}^3}{e^{\frac{hc\tilde{\nu}}{k_{\mathrm{B}}T}} - 1}, \tag{2}$$

were $h$ is the Planck constant and $c$ is the speed of light.

According to Eq. 1, the total energy density of the CMB radiation in the wavenumber domain received in the frequency interval from $\tilde{\nu}_1$ to $\tilde{\nu}_2$ has the following structure:

$$B_0(\tilde{\nu}_1, \tilde{\nu}_2, T) = \frac{4\pi}{c} \left\{ \int_{\tilde{\nu}_1}^{\tilde{\nu}_2} B_{\tilde{\nu}}(T_0) \mathrm{d}\tilde{\nu} + \Delta T \int_{\tilde{\nu}_1}^{\tilde{\nu}_2} \frac{\partial B_{\tilde{\nu}}(T)}{\partial T}\Big|_{T=T_0} \mathrm{d}\tilde{\nu} \right\} \tag{3}$$

Let us note that in (Mather et al. 1994), the variable $\tilde{\nu} = \nu c$ calls as the frequency with the unit [$\mathrm{cm}^{-1}$], where $\nu$ is the frequency [Hz]. However, according to

(http://en.wikipedia.org/wiki/Wavenumber), the variable $\tilde{\nu}$ is called as wavenumber. Further, we will call $\tilde{\nu}$ as the wavenumber and $\nu$ as the frequency.

For constructing the thermodynamics of the CMB radiation, using the *COBE* FIRAS instrument data in the finite range of wavenumbers, hereinafter, it is convenient to present the Planck function Eq. 2 in the frequency domain. Using the relationship

$$B_{\tilde{\nu}}(T)d\tilde{\nu} = B_{\nu(\tilde{\nu})}(T)d\nu, \tag{4}$$

where $\tilde{\nu}$ stands for wavenumber, that can be related to the frequency $\nu$ via the transformation $\nu(\tilde{\nu})$, then

$$B_{\nu}(T) = B_{\tilde{\nu}}(T)\frac{d\tilde{\nu}}{d\nu} = \frac{2h}{c^2}\frac{\nu^3}{e^{\frac{h\nu}{k_B T}}-1} \quad . \tag{5}$$

Eq. 5 is called the Planck functions in the frequency domain.

Using the following relationship between the spectral energy density $I_{\nu}(T)$ and Eq. 5

$$I_{\nu}(T) = \frac{4\pi}{c}B_{\nu}(T), \tag{6}$$

we obtain

$$I_{\nu}(T) = \frac{8\pi h}{c^3}\frac{\nu^3}{e^{\frac{h\nu}{k_B T}}-1} \quad . \tag{7}$$

Then, according to Eq. 3, the total energy density of the CMB radiation in the frequency domain has the form

$$I_0(\nu_1,\nu_2,T) = \int_{\nu_1}^{\nu_2} I_{\nu}(T)\mathrm{d}\nu + \Delta T\int_{\nu_1}^{\nu_2}\frac{\partial I_{\nu}(T)}{\partial T}\Big|_{T=T_0}\mathrm{d}\nu \, . \tag{8}$$

Here the first term is the total energy density for the monopole spectrum and the second one for the dipole spectrum.

Knowledge of the total energy density received in the finite range of frequencies allows us to construct the thermodynamics of the CMB radiation as follows (Landau & Lifshitz 1980):

1) Helmholtz free energy density $f = \frac{F}{V}$:

$$f = -\frac{1}{3}I_0(\nu_1,\nu_2,T) \tag{9}$$

2) Entropy density $s = \frac{S}{V}$:

$$s = \frac{1}{3}\frac{\partial I_0(v_1, v_2, T)}{\partial T} \tag{10}$$

3) Heat capacity at constant volume per unit volume $c_V = \frac{C_V}{V}$:

$$c_V = \frac{T}{3}\left(\frac{\partial^2 I_0(v_1, v_2, T)}{\partial^2 (T)}\right)_V \tag{11}$$

4) Pressure of photons *P*:

$$P = \frac{1}{3} I_0(v_1, v_2, T) \tag{12}$$

5) Enthalpy density $h = \frac{H}{V}$:

$$h = \frac{T}{3}\frac{\partial I_0(v_1, v_2, T)}{\partial T} \tag{13}$$

6) Internal energy density $u = \frac{U}{V}$:

$$u = -\frac{1}{3}\left(I_0(v_1, v_2, T) - T\frac{\partial I_0(v_1, v_2, T)}{\partial T}\right) \tag{14}$$

7) The number density of photons $n = \frac{N}{V}$:

$$n = \frac{8\pi h}{c^3}\int_{v_1}^{v_2} \frac{v^2}{e^x - 1} dv \tag{15}$$

Here *T* is the temperature and *V* is the volume of emitted body.

## 3. MONOPOLE SPECTRUM

Now let us construct the thermodynamics of the CMB radiation for the monopole spectrum. In this case, the total energy density of the CMB radiation received in the finite range of frequencies from $v_1$ to $v_2$ is described by the first term in Eq. 8. Substituting Eq. 7 in Eq. 8 and after computing the integral, we obtain

$$I_0^{\mathrm{M}}(x_1, x_2, T) = \int_{\nu_1}^{\nu_2} I_\nu(T)\mathrm{d}\nu = \frac{8\pi h}{c^3}\int_{\nu_1}^{\nu_2} \frac{\nu^3}{e^{\frac{h\nu}{k_{\mathrm{B}}T}} - 1} d\nu = \frac{48\pi(k_{\mathrm{B}}T)^4}{c^3h^3}[P_3(x_1) - P_3(x_2)] \ . \tag{16}$$

Here $x = \dfrac{h\nu}{k_{\mathrm{B}}T}$. $P_3(x)$ is defined as

$$P_3(x) = \sum_{s=0}^{3} \frac{(x)^s}{s!}\mathrm{Li}_{4-s}(e^{-x}) \, , \tag{17}$$

where

$$\mathrm{Li}_{4-s}(e^{-x}) = \sum_{k=1}^{\infty} \frac{e^{-kx}}{k^{4-s}} \tag{18}$$

is the polylogarithm function (Abramowitz & Stegun 1972).

In the semi-infinite range ($0 \le \nu \le \infty$), Eq. 16 can be re-written as

$$I_0^{\mathrm{M}}(0, \infty, T) = \frac{48\pi k_B{}^4}{c^3h^3} T^4 (P_3(0) - P_3(\infty)) \, . \tag{19}$$

Since $P_3(0) = \mathrm{Li}_4(1) = \xi(4) = \dfrac{\pi^4}{90}$ and $P_3(\infty) = 0$, Eq. 19 transforms to the well-known expression for the total energy density of the blackbody radiation (Landau & Lifshitz 1980)

$$I_0^{\mathrm{M}}(0, \infty, T) = a\,T^4 \, . \tag{20}$$

Here $a$ is the radiation density constant

$$a = \frac{24\pi^5 k_B{}^4}{45c^3h^3} \tag{21}$$

The value of $a$ is $a = 7.5657 \times 10^{-16} \dfrac{\mathrm{J}}{\mathrm{m^3K^4}}$. The Stefan-Boltzmann constant $\sigma$ can be determined using the following relationship $\sigma = \dfrac{ac}{4}$ and, in accordance with Eq. 21, takes the value $\sigma = 5.67037 \times 10^{-8} \dfrac{\mathrm{W}}{\mathrm{m^2K^4}}$ (Landau & Lifshitz 1980). Then the Stefan-Boltzmann law or the total radiation power per unit area received in the semi-infinite frequency range has the well-known form

$$I_0'^{\mathrm{M}}(0, \infty, T) = \sigma\,T^4 \, . \tag{22}$$

Let us note that according to Eq. 22, the total radiation power $I_{\text{total}}^{\text{M}}(T)$ emitted from the area $A$ of early universe, which is presented in the form of a spherical shell of finite thickness at distance of almost 15 billion light years, is defined as follows

$$I_{\text{total}}^{\text{M}}(T) = AI_{\text{o}}'^{\text{M}}(0,\infty,T). \qquad (23)$$

Here it is important to note that the characteristic area $A$ may be obtained from experiment data by measuring the total radiation power $I_{\text{Measured}}^{\text{M}}(T) = I_{\text{total}}^{\text{M}}(T)$, and then using the Eq. 23. In this case, we have

$$A = \frac{I_{\text{Measured}}^{\text{M}}(T)}{I_{\text{o}}'^{\text{M}}(0,\infty,T)} \quad . \qquad (24)$$

For this purpose the optical device for measuring the total radiation power emitted from the area A of a surface at low temperature will be useful.

Let us represent Eq. 16 as follows

$$I_0^{\text{M}}(x_1,x_2,T) = a'(x_1,x_2)T^4, \qquad (25)$$

where

$$a'(x_1,x_2) = \frac{48\pi k_{\text{B}}^{4}}{c^3h^3}[P_3(x_1) - P_3(x_2)]. \qquad (26)$$

According to Eq. 20, $a'(x_1,x_2)$ can be called as the radiation density constant in the finite range of frequencies $v_1 \le v \le v_2$.

Then, the Stefan-Boltzmann law has the following form

$$I_0'^{\text{M}}(x_1,x_2,T) = \sigma'(x_1,x_2)T^4. \qquad (27)$$

Here

$$\sigma'(x_1,x_2) = \frac{a'(x_1,x_2)c}{4} = \frac{12\pi k_{\text{B}}^{4}}{c^2h^3}[P_3(x_1) - P_3(x_2)]. \qquad (28)$$

$\sigma'(x_1,x_2)$, as in the case of the radiation density constant, can be called as the Stefan-Boltzmann constant in the finite range of frequencies.

It is important to note that Eq. 24 has the same structure within a finite range of frequencies in which the values $I_{\text{Measured}}^{\text{M}}(T)$ and $I_0'^{\text{M}}(0,\infty,T)$ have to be replaced by $I_{\text{Measured}}^{\text{M}}(x_1,x_2,T)$ and $I_0'^{\text{M}}(x_1,x_2,T)$.

In accordance with Eq. 13, the number density of photons of the CMB radiation with a photon energy from $h\nu_1$ to $h\nu_2$, defined as follows

$$n = \frac{8\pi h}{c^3}\int_{x_1}^{x_2} \frac{\nu^2}{e^x - 1} d\nu = \frac{16\pi k_{\mathrm{B}}^3}{c^3 h^3} T^3 [P_2(x_1) - P_2(x_2)]. \tag{29}$$

In the case of the semi-infinite range of frequencies, Eq. 29 simplified

$$n = \frac{8\pi h}{c^3}\int_{0}^{\infty} \frac{\nu^2}{e^x - 1} d\nu = \frac{16\pi k_{\mathrm{B}}^3}{c^3 h^3} T^3 [P_2(0) - P_2(\infty)]. \tag{30}$$

Since $P_2(0) = L_{i_3}(1) = \xi(3)$ and $P_2(\infty) = 0$, Eq. 30 transforms to the well-known expression (Landau & Lifshitz 1980)

$$n = \frac{8\pi h}{c^3}\int_{0}^{\infty} \frac{\nu^2}{e^x - 1} d\nu = \frac{2\xi(3)}{\pi^2}\left(\frac{2\pi k_{\mathrm{B}} T}{hc}\right)^3 \approx 0.244\left(\frac{2\pi k_{\mathrm{B}} T}{hc}\right)^3 . \tag{31}$$

According to Eqs, 7 – 13, the thermodynamic function of the CMB radiation for the monopole spectrum in the finite frequency range have the following structure:

(1) Helmholtz free energy density *f*:

$$f = -\frac{16\pi k_B{}^4}{c^3 h^3} T^4 (P_3(x_1) - P_3(x_2)) \tag{32}$$

(2) Entropy density *s*:

$$s = \frac{64\pi k_B{}^4}{c^3 h^3} T^3 \left[ P_3(x_1) - P_3(x_2) + \frac{1}{24}(x_1^4 \operatorname{Li}_0(e^{-x_1}) - x_2^4 \operatorname{Li}_0(e^{-x_2})) \right] \tag{33}$$

(3) Heat capacity at constant volume per unit volume, $c_V$

$$c_V = \frac{192\pi k_B{}^4}{c^3 h^3} T^3 \times$$

$$\left[ P_3(x_1) - P_3(x_2) + \frac{1}{24}(x_1^4 \operatorname{Li}_0(e^{-x_1}) - x_2^4 \operatorname{Li}_0(e^{-x_2})) + \frac{1}{72}(x_1^5 \operatorname{Li}_{-1}(e^{-x_1}) - x_2^5 \operatorname{Li}_{-1}(e^{-x_2}) \right] \tag{34}$$

(4) Pressure *P*:

$$P = \frac{16\pi k_B{}^4}{c^3 h^3} T^4 (P_3(x_1) - P_3(x_2)) \tag{35}$$

(5) Enthalpy density $h$:

$$h = \frac{64\pi k_B{}^4}{c^3 h^3} T^4 \left[ P_3(x_1) - P_3(x_2) + \frac{1}{24}(x_1^4 \operatorname{Li}_0(e^{-x_1}) - x_2^4 \operatorname{Li}_0(e^{-x_2})) \right] \tag{36}$$

(6) Internal energy density $u$:

$$u = \frac{80\pi k_B{}^4}{c^3 h^3} T^4 \left[ (P_3(x_1) - P_3(x_2)) + \frac{1}{30}(x_1^4 \operatorname{Li}_0(e^{-x_1}) - x_2^4 \operatorname{Li}_0(e^{-x_2})) \right] \quad (37)$$

It is not difficult to show that Eqs. 32-37 in the semi-infinite range of frequencies are converted to the well-known expressions for the thermodynamic functions (Landau & Lifshitz 1980).

By definition (Landau & Lifshitz 1980), the Gibbs free energy $G = H - TS$, as clearly seen from Eq. 31 and Eq. 34, is equal to zero. Using the equation $G = \mu N = 0$, we can conclude that the chemical potential $\mu$ is zero, too. Here, $N \neq 0$ is the total number of photons in a system.

Now let us apply obtained expressions for calculating the radiative and thermodynamic functions of the CMB radiation for the monopole spectrum at the temperature $T$ = 2.728 K (Mather et al. 1994; Mather et al. 1997). Using the data obtained by the *COBE* FIRAS instrument in the $60\text{-}600\,\text{GHz}$ frequency interval, the radiative and thermodynamic properties of the CMB radiation are calculated and presented in Table 1. As seen in Table 1, the calculated values in the finite range slightly differ from the corresponding values for the semi-infinite range. For example, the radiation density constant in the range from $\nu_1 = 60$ GHz to $\nu_2 = 600$ GHz is 95% from the corresponding value for the semi-infinite interval. As for entropy density, we have 97%. It means that observed part of spectrum from $\nu_1 = 60$ GHz to $\nu_2 = 600$ GHz covers a significant portion of the total spectrum.

One of the interesting questions is the following. What contribution to the radiative and thermodynamic properties of the CMB radiation gives the Wien part of the total spectrum? The obtained above expressions allow us to answer this question. Indeed, if we assume that the observed results suggest that the Rayleigh-Jeans region is described by the Planck function, we have to calculate the radiative and thermodynamic functions of the CMB radiation in the range of $0 \leq \nu \leq 60\,\text{GHz}$. Then the radiative and thermodynamic properties in Wien's region $600\,\text{GHz} \leq \nu \leq \infty$ can be calculated by subtracting the Rayleigh-Jeans region $0 \leq \nu \leq 60\,\text{GHz}$ and the Planck part of the spectrum $60\text{ GHz} \leq \nu_1 \leq 600\text{ GHz}$ from the total spectrum in the semi-infinite frequency range. Performing the calculation, for the Wien part of the spectrum $600\,\text{GHz} \leq \nu \leq \infty$, we obtain $a = 4.8203 \times 10^{-18} \frac{\text{J}}{\text{m}^3\text{K}^4}$ and $s = 3.8661 \times 10^{-16} \frac{\text{J}}{\text{m}^3\text{K}}$. Then, the contribution of Wien's part of the spectrum to the radiation density constant $a$ is 0.64% from the corresponding value for the semi-infinite interval. As for the contribution to entropy density, we have 1.79%. As you can see, the Wien part of a total spectrum gives a small contribution to the radiative and thermodynamic properties of CMB radiation.

## 4. DIPOLE SPECTRUM

The first anisotropy discovered was the dipole anisotropy. The dipole spectrum of the CMB radiation is generally interpreted as a Doppler shift due to the Earth's motion relative to the CMB radiation field. The *Cobe* Firas instrument was used for the measurement of the dipole spectrum in the wavenumber range between 2 and 20 cm$^{-1}$ (Fixsen, D.J. et al. 1994; Fixsen, D.J. et al. 1996). The observed dipole spectrum, a second term in Eq. 1, was fitted by the following expression

$$I^{D}(\tilde{\nu},T)=\frac{I_{0}(\tilde{\nu})}{\left(\frac{\upsilon}{c}\right)\cos(\theta)}=T_{\mathrm{amp}}\left.\frac{\partial B_{\tilde{\nu}}(T)}{\partial T}\right|_{T=T_{0}} . \tag{38}$$

Here $T_0 = 2.728\ \mathrm{K}$ . The best-fit value of the dipole amplitude is $T_{\mathrm{amp}} = 3.369 \pm 0.004\ \mathrm{mK}$ .

According to Eq. 3, the total energy density is defined as

$$I^{D}(\tilde{\nu}_1,\tilde{\nu}_2,T)=\frac{4\pi}{c}T_{\mathrm{amp}}\int_{\tilde{\nu}_1}^{\tilde{\nu}_2}\frac{\partial B_{\tilde{\nu}}(T)}{\partial T}d\tilde{\nu}=\frac{4\pi}{c}T_{\mathrm{amp}}\frac{\partial}{\partial T}\int_{\tilde{\nu}_1}^{\tilde{\nu}_2}B_{\tilde{\nu}}(T)d\tilde{\nu} . \tag{39}$$

Using the similar procedure for the transition to the frequency domain, as in the case of the monopole spectrum, the expression for the total energy density of the CMB radiation in the finite range of frequencies for the dipole spectrum has the following form:

$$I_0^{D}(x_1,x_2,T)=T_{\mathrm{amp}}\frac{\partial}{\partial T}\int_{\nu_1}^{\nu_2}I_{\nu}(T)\mathrm{d}\nu=T_{\mathrm{amp}}\frac{48\pi k_{\mathrm{B}}}{c^3h^3}\frac{\partial}{\partial T}\left(T^4[P_3(x_1)-P_3(x_2)]\right). \tag{40}$$

After differentiating, Eq. (40) is presented as

$$I_0^{D}(x_1,x_2,T)=T_{\mathrm{amp}}\frac{192\pi k_B{}^4}{c^3h^3}T_0{}^3\left[P_3(x_1)-P_3(x_2)+\frac{1}{24}(x_1^4\ \mathrm{Li}_0(e^{-x_1})-x_2^4\ \mathrm{Li}_0(e^{-x_2}))\right]. \tag{41}$$

Since $P_3(0)=\mathrm{Li}_4(1)=\xi(4)=\frac{\pi^4}{90}$ and $P_3(\infty)=0$, and $x_1^4\ \mathrm{Li}_0(e^{-x_1})\big|_{x_1=0}=x_2^4\ \mathrm{Li}_0(e^{-x_2})\big|_{x_2=\infty}=0$, the total energy density of the CMB radiation in the semi-infinite range of frequencies has the form

$$I_0^D(0,\infty,T) = \frac{32\pi^5 k_B{}^4}{15c^3h^3} T_{\text{amp}} T_0{}^3 . \quad (42)$$

Here

$$a'' = \frac{32\pi^5 k_B{}^4}{15c^3h^3} \quad (43)$$

can be called as the radiation density constant for the dipole spectrum in the semi-infinite range of frequencies. The value of $a''$ is $a'' = 3.0263\times10^{-15} \frac{\text{J}}{\text{m}^3\text{K}^4}$. The Stefan-Boltzmann constant for the dipole spectrum is determined by the following relationship $\sigma'' = \frac{a''c}{4}$. Then, $\sigma''$ has the following structure:

$$\sigma'' = \frac{8\pi^5 k_B{}^4}{15c^2h^3} . \quad (44)$$

The value of $\sigma''$ is $\sigma'' = 2.2681\times10^{-7}\ \text{J m}^{-2}\ \text{s}^{-1}\ \text{K}^{-4}$. Then, the Stefan-Boltzmann law or the total radiation power per unit area $I_0'^D(T)$ for the dipole spectrum is defined as

$$I_0'^D(0,\infty,T) = \frac{8\pi^5 k_B{}^4}{15c^2h^3} T_{\text{amp}} T_0{}^3 . \quad (45)$$

According to Eq. 45, the total radiation power emitted from the area $A'$ of the universe for the dipole spectrum can be determined as follows

$$I'^{\text{D}}_{\text{total}}(T) = A' I'^{\text{D}}_{\text{o}}(0,\infty,T) . \quad (46)$$

Therefore, as in the case of the monopole spectrum, the characteristic area $A'$ may be obtained from the experiment data by the measurement of the total radiation power $I^{\text{D}}_{\text{Measured}}(v_1, v_2, T)$. Then, using Eq. 46, we have

$$A' = \frac{I^{\text{D}}_{\text{Measured}}(T)}{I'^{\text{D}}_{\text{o}}(0,\infty,T)} . \quad (47)$$

Here it is important to note that the radiation density constant and the Stefan-Boltzmann constant for the dipole spectrum, Eq. 43 and Eq. 44, in the semi-infinite range of frequencies differ from the corresponding constants for the monopole spectrum. See Table 1. It means that Doppler shift leads to a renormalization of the corresponding constants. This situation is similar for the radiative and thermodynamic functions of the CNB radiation for the dipole spectrum in a semi-infinite range.

Let us Eq. 41 present in the form

$$I_0^D(x_1, x_2, T) = a''(x_1, x_2) T_{\mathrm{amp}} T_0^{\,3} \quad , \tag{48}$$

where

$$a''(x_1, x_2) = \frac{192\pi k_B^{\,4}}{c^3 h^3}\left[ P_3(x_1) - P_3(x_2) + \frac{1}{24}(x_1^4 \,\mathrm{Li}_0(e^{-x_1}) - x_2^4 \,\mathrm{Li}_0(e^{-x_2})) \right] \tag{49}$$

can be termed as the radiation density constant for the dipole spectrum in the finite range of frequencies. The Stefan-Boltzmann constant $\sigma''(v_1, v_2, T)$ is defined as

$$\sigma''(x_1, x_2) = \frac{4\pi a''(x_1, x_2)}{c} = \frac{48\pi k_B^{\,4}}{c^2 h^3}\left[ P_3(x_1) - P_3(x_2) + \frac{1}{24}(x_1^4 \,\mathrm{Li}_0(e^{-x_1}) - x_2^4 \,\mathrm{Li}_0(e^{-x_2})) \right] \tag{50}$$

Then the Stefan-Boltzmann law or the total radiation power per unit area in the finite range of frequencies for the dipole spectrum has the structure

$$I_0'^D(x_1, x_2, T) = \sigma''(x_1, x_2)\, T_{\mathrm{amp}} T_0^{\,3} \,. \tag{51}$$

According to Eq. 15, the number density of photons with photon energy from $hv_1$ to $hv_2$ is defined as follows

$$n = \frac{16\pi k_{\mathrm{B}}^3}{c^3 h^3} T_{\mathrm{amp}} \frac{\partial}{\partial T}\left[ T^3 (P_2(x_1) - P_2(x_2)) \right] . \tag{52}$$

After differentiating, Eq. 52 takes the form

$$n = \frac{48\pi k_{\mathrm{B}}^3}{c^3 h^3} T_{\mathrm{amp}} T^2 \{ [(P_2(x_1) - P_2(x_2)] + \frac{1}{6}(x_1^2 \,\mathrm{Li}_1(e^{-x_1}) - x_2^2 \,\mathrm{Li}_1(e^{-x_{21}})) \} . \tag{53}$$

In the case of the semi-infinite range $0 \le v \le \infty$ , Eq. 53 is simplified as

$$n = \frac{48\pi k_{\mathrm{B}}^3}{c^3 h^3} T_{\mathrm{amp}} T^2 [(P_2(0) - P_2(\infty)] \,. \tag{54}$$

Since $P_2(0) = L_{i_3}(1) = \varsigma(3)$ and $P_2(\infty) = 0$ , Eq. 39 is defined as

$$n = \frac{57.696\,\pi k_{\mathrm{B}}^3}{c^3 h^3} T_{\mathrm{amp}} T^2 \quad . \tag{55}$$

Using Eqs.7-13 and Eq. 41, the thermodynamic functions of the CMB radiation in the finite $v_1 \le v \le v_2$ and the semi-infinite $0 \le v \le \infty$ ranges of frequencies for the dipole spectrum have the following structure:

(1) Helmholtz free energy density *f*:

a) $v_1 \le v \le v_2$

$$f = -\frac{64\pi k_B{}^4}{c^3 h^3} T_{\text{amp}} T_0{}^3 \left[ P_3(x_1) - P_3(x_2) + \frac{1}{24}(x_1^4 \operatorname{Li}_0(e^{-x_1}) - x_2^4 \operatorname{Li}_0(e^{-x_2})) \right] \tag{56}$$

b) $0 \le v \le \infty$

$$f = -\frac{32\pi^5 k_B{}^4}{45 c^3 h^3} T_{\text{amp}} T_0{}^3 \tag{57}$$

(2) Entropy density *s*:

a) $v_1 \le v \le v_2$

$$s = \frac{192\pi k_B{}^4}{c^3 h^3} T_{\text{amp}} T^2 \times$$

$$\left[ P_3(x_1) - P_3(x_2) + \frac{1}{24}(x_1^4 \operatorname{Li}_0(e^{-x_1}) - x_2^4 \operatorname{Li}_0(e^{-x_2})) + \frac{1}{72}(x_1^5 \operatorname{Li}_{-1}(e^{-x_1}) - x_2^5 \operatorname{Li}_{-1}(e^{-x_2}) \right] \tag{58}$$

b) $0 \le v \le \infty$

$$s = \frac{32\pi^5 k_B{}^4}{15 c^3 h^3} T_{\text{amp}} T^2 \tag{59}$$

(3) Heat capacity at constant volume per unit volume $c_V$ :

a) $v_1 \le v \le v_2$

$$c_V = \frac{384\pi k_B{}^4}{c^3 h^3} T_{\text{amp}} T^2 \times$$

$$\left[ P_3(x_1) - P_3(x_2) + \frac{1}{24}(x_1^4 \operatorname{Li}_0(e^{-x_1}) - x_2^4 \operatorname{Li}_0(e^{-x_2})) + \frac{1}{144}(x_1^6 \operatorname{Li}_{-2}(e^{-x_1}) - x_2^6 \operatorname{Li}_{-2}(e^{-x_2}) \right] \tag{60}$$

b) $0 \le v \le \infty$

$$c_V = \frac{32\pi^5 k_B{}^4}{15 c^3 h^3} T_{\text{amp}} T^2 \tag{61}$$

(4) Pressure *P*:

a) $v_1 \le v \le v_2$

$$P = \frac{64\pi k_B{}^4}{c^3 h^3} T_{\text{amp}} T_0{}^3 \left[ P_3(x_1) - P_3(x_2) + \frac{1}{24}(x_1^4 \operatorname{Li}_0(e^{-x_1}) - x_2^4 \operatorname{Li}_0(e^{-x_2})) \right] \tag{62}$$

b) $0 \le v \le \infty$

$$P=\frac{32\pi^5 k_B{}^4}{45c^3h^3}T_{\mathrm{amp}}T_0{}^3 \tag{63}$$

(5) Enthalpy density *h*:

a) $\nu_1 \le \nu \le \nu_2$

$$h=\frac{192\pi k_B{}^4}{c^3h^3}T_{\mathrm{amp}}T^3\times$$

$$\left[P_3(x_1)-P_3(x_2)+\frac{1}{24}(x_1^4\,\mathrm{Li}_0(e^{-x_1})-x_2^4\,\mathrm{Li}_0(e^{-x_2}))+\frac{1}{72}(x_1^5\,\mathrm{Li}_{-1}(e^{-x_1})-x_2^5\,\mathrm{Li}_{-1}(e^{-x_2})\right] \tag{64}$$

b) $0 \le \nu \le \infty$

$$h=\frac{32\pi^5 k_B{}^4}{15c^3h^3}T_{\mathrm{amp}}T^3 \tag{65}$$

(6) Internal energy density *u*:

a) $\nu_1 \le \nu \le \nu_2$

$$u=\frac{384\pi k_B{}^4}{c^3h^3}T_{\mathrm{amp}}T^3\times$$

$$\left[P_3(x_1)-P_3(x_2)+\frac{1}{24}(x_1^4\,\mathrm{Li}_0(e^{-x_1})-x_2^4\,\mathrm{Li}_0(e^{-x_2}))+\frac{1}{48}(x_1^5\,\mathrm{Li}_{-1}(e^{-x_1})-x_2^5\,\mathrm{Li}_{-1}(e^{-x_2})\right] \tag{66}$$

b) $0 \le \nu \le \infty$

$$u=\frac{32\pi^5 k_B{}^4}{15c^3h^3}T_{\mathrm{amp}}T^3 \tag{67}$$

Gibbs free energy $G = H - TS$, as clearly seen from Eqs. 58, 64 and Eqs. 59, 65, is equal to zero. Since $G = \mu N = 0$ and the total number of photons $N \ne 0$ implies that the chemical potential $\mu$ is zero, too.

Now let us calculate the thermodynamic and radiative properties of the CMB radiation for the dipole spectrum using *Cobe* FIRAS instrument data (Fixsen, D.J. et al. 1994; Fixsen, D.J. et al. 1996). In Table 1, the values for the thermodynamic and radiative functions for the dipole spectrum in the finite and semi-infinite ranges of frequencies at $T$ = 2.728 K and $T_{\mathrm{amp}}$ = 3.369 mK are presented. As can be seen in Table 1, the radiation density constant $a''$ and the Stefan-Boltzmann constant $\sigma''$ for the dipole spectrum differ from those of the constants $a'$ and $\sigma'$ for the monopole spectrum. This means that Doppler shift leads to a renormalization of these constants, as well as the corresponding constants for the thermodynamic and radiative functions.

It is important to note that the exact expressions for the radiative and thermodynamic functions, obtained above, can be used to construct the thermodynamics of the CMB radiation for the quadrupole and a higher order of spectra. Indeed, in the case of the quadrupole spectrum, for example, we need to include the quadrupole term $I^{Q}(\tilde{v},T)=\frac{1}{2}(\Delta T)^{2}\left.\frac{\partial^{2}B_{\tilde{v}}(T)}{\partial T^{2}}\right|_{T=T_0}$ in Eq. 1. Using Eq. 41, $\left.\frac{\partial^{2}B_{\tilde{v}}(T)}{\partial T^{2}}\right|_{T=T_0}$ has the form $\left.\frac{\partial^{2}B_{\tilde{v}}(T)}{\partial T^{2}}\right|_{T=T_0}=\left.\frac{\partial I_0^{D}(v_1,v_2,T)}{\partial T}\right|_{T-T_0}$. After differentiating, the thermodynamics of the CMB radiation (Eq. 7 – 11) for the quadrupole spectrum will be constructed and the radiative properties will be determined. New renormalized values of the radiation density $a$ and the Stefan-Boltzmann $\sigma$ will be obtained. As a result, the contributions of the quadrupole and a higher order of spectra to the total spectrum of the CMB radiation will be obtained.

In conclusion, it should be noted that the Cosmic Microwave Background radiation is the largest observed redshift, which corresponds to the greatest distance. According to the Cosmic detectives (2013), this value of the redshift is around $z$ = 1089 and it shows the state of the Universe about 13.8 billion years ago. Now we calculate the radiative and thermodynamic functions for the monopole and dipole spectra in the finite and semi-infinite ranges of frequencies at the redshift $z$ = 1089. It is well known that in an expanding universe the temperature $T$ and the frequency $v$ depends on the redshift z (Sunyaev & Zel'dovich 1980), in accordance with the formulas $v=v_0(1+z)$ and $T=T_0(1+z)$. In this case, the obtained expressions for the radiative and thermodynamic functions of the CMB radiation for the monopole and dipole spectra have the same structure, in which the temperatures $T$ = 2.728 K and $T_{\text{amp}}=3.369$ mK should be replaced by $T$ = 2973.54 K and $T_{\text{amp}}=3.67\,\text{K}$. Since at the redshift $z$ = 1089, the temperature $T$ by three orders of magnitude greater than the temperature of CMB radiation at present time $z$ = 0, let us, for example, conduct the calculation using a frequency range from $v_1=4.99\times10^{13}\,\text{Hz}$ and $v_2=4.99\times10^{14}\,\text{Hz}$. This range belongs to the Planck part of the total spectrum. Indeed, in accordance with the Win displacement law (Landau & Lifshitz 1980), the position of the maximum of the spectral energy density is determined as $\nu_{\max} = 2.8214\dfrac{k_{\mathrm{B}}T}{h}$ and equal to $\nu_{\max} = 6.1959 \times 10^{13}\,\mathrm{Hz}$. This value of $\nu_{\max}$ is within the selected range of frequencies. In Table 2, the radiative and the thermodynamic functions for the monopole and dipole spectra in the finite and semi-infinite frequency ranges at $z$ = 1089 are presented.

## 5. SUMMARY AND CONCLUSIONS

In this paper, the exact expressions for the calculation of the temperature dependences of the radiative and thermodynamic functions of the Cosmic Microwave Background (CMB) radiation, such as the total radiation power per unit area, total energy density, number density of photons, Helmholtz free energy density, entropy density, heat capacity at constant volume, pressure, enthalpy density, and internal energy density in the finite range of frequencies are obtained.

Utilizing the experimental data for the monopole spectrum measured by the *COBE* FIRAS instrument in the finite range of frequencies $60\mathrm{GHz} \leq \nu \leq 600\mathrm{GHz}$ at the temperature $T$ = 2.728 K, the values of the radiative and thermodynamic functions, as well as the radiation density constant $a'$ and the Stefan-Boltzmann constant $\sigma'$ are calculated. For the dipole spectrum, the constants $a''$ and $\sigma''$, and the radiative and thermodynamic properties of CMB radiation are obtained using the mean amplitude $T_{\mathrm{amp}} = 3.369$ mK . The results are presented in Table 1. It is shown that the Doppler shift leads to a renormalization of the radiation density constant, the Stefan-Boltzmann constant, and the corresponding constants for the thermodynamic functions.

Knowing the dependence of the temperature $T$ on the redshift z allows us to study the thermodynamic and radiative state of the Universe many years ago. As an example, the thermodynamic and radiative functions, which belong to the state of the Universe at redshift $z$ = 1089, corresponding to the state of Universe about 13.8 billion years ago, for the monopole and dipole spectra in the semi-infinite $0 \leq \nu \leq \infty$ and finite $0.0499\,\mathrm{PHz} \leq \mathrm{v} \leq 0.499\,\mathrm{PHz}$ ranges of frequencies are calculated. The calculated values are presented in Table 2. These values differ significantly from corresponding values at present time $z$=0.

In conclusion, it is important to note the following directions for the future research:

a) It is desirable to investigate the contributions of the quadrupole and a higher order of spectra to the total spectrum of the CMB radiation (Eq. 1). In this case, we obtain new temperature dependencies of the radiative and thermodynamic properties of the CMB radiation.

b) One of the important issues is to construct the thermodynamic and radiative functions of Galactic radiation using the Galaxy spectrum. The latter has the form $v^n B_v(T_{dust})$ . The index *n* changes from 1.65 to 2. Fixen et al (1996) examined the FIRAS Galaxy spectrum and found that it was fitted using *n* = 2 with $T_{dust} = 13 \pm 1\,\mathrm{K}$. As a result, the exact expressions for the thermodynamic and radiative properties of the Galactic radiation can be obtained.

c) Particular attention should be paid to the investigation of the radiative and thermal properties of the extragalactic Far Infrared Background (FIRB) radiation (Fixen et al. 1998). In this case, the radiative and thermodynamic functions of the FIRB radiation in the frequency interval $\tilde{v} = 5 - 80\,\mathrm{cm}^{-1}$ at $T = 18.5 \pm 1.2\,\mathrm{K}$ will be defined.

d) In the future, it is of interest in the study of the temperature *T* and the redshift *z* dependences of the Wien displacement law for the dipole spectrum of the CMB radiation. As a result, different law of the relationships between the position of the maximum of the spectral energy density and temperature will be established (Fisenko & Ivashov 1999).

e) Developed in this work approach can be useful to consider the Hawking radiation. This radiation is predicted to be released by black holes that emit exactly blackbody radiation.

These and other topics will be points of discussion in subsequent publication.

**Acknowledgments**

The authors sincerely thank Dr. J.C. Mather for the consultations.

## REFFERENCES

| **Quantity** | **Monopole** $0 \le \tilde{\nu} \le \infty$ | **Monopole** $\nu_1 \le \nu \le \nu_2$ | **Dipole** $\nu_1 \le \nu \le \nu_2$ | **Dipole** $0 \le \tilde{\nu} \le \infty$ |
|---|---|---|---|---|
| $a'$, $a''$ $\left[\mathrm{J\ m^{-3}\ K^{-4}}\right]$ | $7.5657\times10^{-16}$ | $7.2170\times10^{-16}$ | $2.9260\times10^{-15}$ | $3.0263\times10^{-15}$ |
| $\sigma'$, $\sigma''$ $\left[\mathrm{W\ m^{-3}\ K^{-4}}\right]$ | $5.6704\times10^{-8}$ | $5.4090\times10^{-8}$ | $2.1930\times10^{-7}$ | $2.2681\times10^{-7}$ |
| $I_0^{\mathrm{M}}(\nu_1,\nu_2,T)$ <br> $I_0^{\mathrm{D}}(\nu_1,\nu_2,T)$ <br> $\left[\mathrm{J\ m^{-3}}\right]$ | $4.1902\times10^{-14}$ <br> - | $3.9970\times10^{-14}$ <br> - | - <br> $2.0013\times10^{-16}$ | - <br> $2.0698\times10^{-16}$ |
| $I_0'^{\mathrm{M}}(\nu_1,\nu_2,T)$ <br> $I_0'^{\mathrm{D}}(\nu_1,\nu_2,T)$ <br> $\left[\mathrm{W\ m^{-2}}\right]$ | $3.1404\times10^{-6}$ <br> - | $2.9956\times10^{-6}$ <br> - | - <br> $1.4999\times10^{-8}$ | - <br> $1.5513\times10^{-8}$ |
| $f$ $\left[\mathrm{J\ m^{-3}}\right]$ | $-1.3967\times10^{-14}$ | $-1.3323\times10^{-14}$ | $-6.6710\times10^{-17}$ | $-6.8993\times10^{-17}$ |
| $u$ $\left[\mathrm{J\ m^{-3}}\right]$ | $6.9836\times10^{-14}$ | $6.7341\times10^{-14}$ | $3.8148\times10^{-16}$ | $4.1397\times10^{-16}$ |
| $h$ $\left[\mathrm{J\ m^{-3}}\right]$ | $5.5869\times10^{-14}$ | $5.4017\times10^{-14}$ | $1.9387\times10^{-16}$ | $2.0698\times10^{-16}$ |
| $s$ $\left[\mathrm{J\ m^{-3}\ K^{-1}}\right]$ | $2.0478\times10^{-14}$ | $1.9801\times10^{-14}$ | $7.1067\times10^{-17}$ | $7.5875\times10^{-17}$ |
| $P$ $\left[\mathrm{N\ m^{-2}}\right]$ | $1.3967\times10^{-14}$ | $1.3323\times10^{-14}$ | $6.6710\times10^{-17}$ | $6.8993\times10^{-17}$ |
| $c_V$ $\left[\mathrm{J\ m^{-3}\ K^{-1}}\right]$ | $6.1439\times10^{-14}$ | $5.7546\times10^{-14}$ | $3.1045\times10^{-16}$ | $4.1397\times10^{-16}$ |
| $n$ | $4.1186\times10^{8}$ | $3.4509\times10^{8}$ | $1.3787\times10^{6}$ | $1.5259\times10^{6}$ |

**Table 1** Calculated values of the radiative and thermodynamic state functions for the monopole and dipole spectra in the 60 – 600 GHz frequency interval at $T$ = 2.728 K and $T_{\mathrm{amp}} = 3.369\ \mathrm{mK}$ .

| Quantity | Monopole $0 \le \tilde{\nu} \le \infty$ | Monopole $\nu_1 \le \nu \le \nu_2$ | Dipole $\nu_1 \le \nu \le \nu_2$ | Dipole $0 \le \tilde{\nu} \le \infty$ |
|---|---|---|---|---|
| $a'$, $a''$ $\left[\mathrm{J\ m^{-3}\ K^{-4}}\right]$ | $7.5657\times10^{-16}$ | $7.1334\times10^{-16}$ | $2.7381\times10^{-15}$ | $3.0263\times10^{-15}$ |
| $\sigma'$, $\sigma''$ $\left[\mathrm{W\ m^{-3}\ K^{-4}}\right]$ | $5.6704\times10^{-8}$ | $5.3464\times10^{-8}$ | $2.0521\times10^{-7}$ | $2.2681\times10^{-7}$ |
| $I_0^{\mathrm{M}}(\nu_1,\nu_2,T)$<br>$I_0^{\mathrm{D}}(\nu_1,\nu_2,T)$<br>$\left[\mathrm{J\ m^{-3}}\right]$ | $5.9149\times10^{-2}$<br>- | $5.5796\times10^{-2}$<br>- | -<br>$2.6419\times10^{-4}$ | -<br>$2.9201\times10^{-4}$ |
| $I_0'^{\mathrm{M}}(\nu_1,\nu_2,T)$<br>$I_0'^{\mathrm{D}}(\nu_1,\nu_2,T)$<br>$\left[\mathrm{W\ m^{-2}}\right]$ | $4.4331\times10^{6}$<br>- | $4.1797\times10^{6}$<br>- | -<br>$1.9801\times10^{4}$ | -<br>$2.1885\times10^{4}$ |
| $f\ \left[\mathrm{J\ m^{-3}}\right]$ | $-1.9716\times10^{-2}$ | $-1.8589\times10^{-2}$ | $-8.8066\times10^{-5}$ | $-9.7337\times10^{-5}$ |
| $u\ \left[\mathrm{J\ m^{-3}}\right]$ | $9.8581\times10^{-2}$ | $8.9943\times10^{-2}$ | $4.1330\times10^{-4}$ | $5.8402\times10^{-4}$ |
| $h\left[\mathrm{J\ m^{-3}}\right]$ | $7.8865\times10^{-2}$ | $7.1354\times10^{-2}$ | $2.2583\times10^{-4}$ | $2.9201\times10^{-4}$ |
| $s\left[\mathrm{J\ m^{-3}\ K^{-1}}\right]$ | $2.6522\times10^{-5}$ | $2.3996\times10^{-5}$ | $7.5947\times10^{-8}$ | $9.8203\times10^{-8}$ |
| $P\left[\mathrm{N\ m^{-2}}\right]$ | $1.9716\times10^{-2}$ | $1.8589\times10^{-2}$ | $8.8066\times10^{-5}$ | $9.7337\times10^{-5}$ |
| $c_V\ \left[\mathrm{J\ m^{-3}\ K^{-1}}\right]$ | $7.9567\times10^{-5}$ | $6.1535\times10^{-5}$ | $2.0770\times10^{-4}$ | $5.8402\times10^{-4}$ |
| $n$ | $5{,}3338\times10^{17}$ | $4.7288\times10^{17}$ | $1.8506\times10^{15}$ | $1.9749\times10^{15}$ |

**Table 2** Calculated values of the radiative and thermodynamic state functions for the monopole and dipole spectra in the 0.0499 – 0.499 PHz frequency interval at the redshift $z \approx 1089$ and at $T = 2973.54\ \mathrm{K}$ and $T_{\mathrm{amp}} = 3.67\ \mathrm{K}$.